\documentclass[aps,prl,superscriptaddress,showpacs,floatfix,nofootinbib,notitlepage,twocolumn]{revtex4-1}
\usepackage{amsmath,graphicx,float,hyperref}

\def\bra{\langle}
\def\ket{\rangle}
\def\bbra{\langle\!\langle}
\def\kket{\rangle\!\rangle}
\newcommand{\trento}{T$\mathrel{\protect\raisebox{-2.1pt}{R}}$ENTo}

\begin{document}

\title{Turning up and down strong magnetic fields in relativistic nuclear collisions}

\author{Giuliano Giacalone}
\affiliation{Universit\'e Paris Saclay, CNRS, CEA, Institut de physique th\'eorique, 91191 Gif-sur-Yvette, France}

\begin{abstract}
I show that the average transverse momentum, $\langle p_t \rangle$, of the hadrons emitted in relativistic nuclear collisions can be used as a ``knob'' to control the strength of the magnetic field induced by the spectator and the participant protons over the overlap region. I thus argue that any observable sensitive to this magnetic field is nontrivially correlated with $\langle p_t \rangle$ at a given collision centrality.
\end{abstract}

\maketitle

Heavy atomic nuclei are smashed at relativistic energy at the BNL Relativistic Heavy Ion Collider (RHIC) and at the CERN Large Hadron Collider (LHC) to produce and characterize the quark-gluon plasma, the hot fluidlike state of strong-interaction matter. These processes involve the interaction of highly charged objects, i.e., ions with $Z\sim80$, moving in opposite directions at nearly the speed of light, and are therefore associated with the emergence of magnetic fields of gigantic strength~\cite{Kharzeev:2007jp,Skokov:2009qp,Bzdak:2011yy,Deng:2012pc,Zhong:2014cda}, $B\sim10^{15}$~T, the strongest ever created in a laboratory.  Experimental searches for signatures of the magnetic field in relativistic nuclear collisions are actively pursued at both RHIC~\cite{Abelev:2009ac,Abelev:2009ad,Adamczyk:2013kcb,Adamczyk:2014mzf,STAR:2019xzd,Adam:2019wnk,STAR:2020crk,Adam:2020zsu} and LHC~\cite{Abelev:2012pa,Khachatryan:2016got,Sirunyan:2017quh,Acharya:2019ijj,Acharya:2020rlz}, and theoretical studies aimed at establishing a quantitative phenomenology of $B$ field-related effects have recently appeared in the literature~\cite{McLerran:2013hla,Roy:2017yvg,Huang:2017tsq,Gursoy:2018yai,Muller:2018ibh,Fukushima:2018grm,Zhao:2019crj,Siddique:2019gqh,Inghirami:2019mkc,Hammelmann:2019vwd,Xu:2020sui,Liang:2020sgr}. This effort is driven by the fact that a strong $B$ field acting on the hot quark-gluon medium may lead to the emergence of so-called chiral anomalous effects~\cite{Kharzeev:2015znc,Landsteiner:2016led,Hattori:2016emy,Zhao:2019hta,Li:2020dwr,Gao:2020vbh,Hou:2020zhb}, whose experimental observation would have far-reaching implications, bringing evidence of local strong parity violation in high-energy nuclear experiments. 

However, this is an outstanding challenge. The observable effects driven by the $B$ field are typically of the same kind as the observable effects driven by the strong interaction governing the quark-gluon plasma~\cite{Tuchin:2013ie}, and it is difficult to separate these two contributions in the data. The strength of the $B$ field in a given collision depends on the number of participant protons, $N_{\rm part}$, and on the number of spectator protons, $N_s$. In this paper, I introduce a new method that allows one to have an experimental handle on these numbers, and thus on the manifestations of the strong $B$ field.

The idea is to look at events that yield the same number of particles in the final state (i.e., same \textit{multiplicity}), and then sort these events according the mean transverse momentum, $\bra p_t \ket$, of their final-state hadrons. In the hydrodynamic framework of high-energy nuclear collisions, the mean transverse momentum is a measure of the energy of the fluid from which the particles are emitted~\cite{Gardim:2019xjs,Gardim:2020sma,Giacalone:2020dln}. In one event, and assuming that the quark-gluon plasma is invariant under longitudinal boosts:
\begin{equation}
\label{eq:barpt}
\bra p_t \ket = \frac{1}{N} \int_{{\bf p}_t} \frac{dN}{d^2{\bf p}_t} p_t,
\end{equation}
where $N$ is the total number of particles detected in one event, and $\frac{dN}{d^2{\bf p}_t}$ is the spectrum of all charged hadrons observed at a given rapidity. Now, collisions with fixed final-state multiplicity correspond to a good approximation to events where the entropy of the medium is fixed. As a consequence, at fixed multiplicity there exists a tight correlation between $\bra p_t \ket$ and the size of the system, because if two events have the same entropy, but different volumes, then the event contained within a smaller volume corresponds to a medium with larger energy, and in turn a larger $\bra p_t \ket$. This well-known feature of hydrodynamics~\cite{Broniowski:2009fm,Mazeliauskas:2015efa,Bozek:2017elk,Schenke:2020uqq,Giacalone:2020dln} implies that a selection of events based on $\bra p_t \ket$ at fixed multiplicity corresponds to a selection based on their size, where a large system size corresponds to a small value of $\bra p_t \ket$, and vice versa.

The argument of the present paper is that the variation of system size induced by a variation of $\bra p_t \ket$ corresponds in turn to a significant variation of the collision impact parameter, and consequently of the number of nucleons that participate, or do not participate, in the collision. The value of $\bra p_t \ket$ provides thus an experimental handle on $N_{\rm part}$ and $N_{\rm s}$, with nontrivial implications for the manifestation of the $B$ field at a given collision centrality.

To show that this works in practice, I perform simulations of the collision process using a phenomenological model. I use the \trento{} model of initial conditions~\cite{Moreland:2014oya}, tuned as in Ref.~\cite{Giacalone:2017dud} to simulate $^{208}$Pb+$^{208}$Pb collisions at LHC. This model provides a prescription for the entropy density, $s({\bf x}, \tau_0)$, created in the interaction of two nuclei $A$ and $B$: $s({\bf x}, \tau_0) \propto \sqrt{T_A ({\bf x}+{\bf b}/2)T_B({\bf x}-{\bf b}/2)}$, where $\tau_0$ is the time at which the hydrodynamics description of the system becomes applicable, ${\bf b}$ is the impact
parameter of the collision, and $T_{A(B)}$ is a Lorentz-boosted density of participant matter.  The \trento{} model does not allow to evaluate $\bra p_t \ket$, nevertheless, following recent studies~\cite{Gardim:2019xjs,Gardim:2020sma}, a good approximation of the relative variation of this quantity can be obtained as follows. I denote the initial energy per rapidity and the initial entropy per rapidity in the quark-gluon plasma respectively by:
\begin{equation}
    \noindent E_0 = \tau_0 \int_{\bf x} e({\bf x},\tau_0), \hspace{30pt} S = \tau_0 \int_{\bf x} s({\bf x},\tau_0),
\end{equation}
where $e({\bf x},\tau_0)$ is the energy density of the system, $e \propto s^{4/3}$, at the beginning of hydrodynamics. Dubbing $\delta\bra p_t \ket = \bra p_t \ket - \bbra p_t \kket $, where $\bbra p_t \kket$ is the average value of $\bra p_t \ket$ at a given centrality, the relative variation of the average transverse momentum is then provided by:
\begin{equation}
\label{eq:predictor}
   \frac{\delta \bra p_t \ket}{\bbra p_t \kket} =  \kappa_0 \frac{E_0/S  - \bra E_0/S \ket}{\bra E_0/S \ket},
\end{equation}
where $\kappa_0$ is a constant which depends on the thermodynamic and viscous properties of the system~\cite{Schenke:2020uqq,Giacalone:2020lbm}, and has to be chosen to reproduce the relative dynamical fluctuation of $\bra p_t \ket$ measured in experimental data~\cite{Abelev:2014ckr,Adam:2019rsf}. Doing so, one can study observables as a function of the relative variation of the final-state $\bra p_t \ket$. I calculate observables at fixed \textit{centrality}. Following the experimental procedure, where the centrality of a collision is defined by the multiplicity~\cite{Abelev:2013qoq,Aaboud:2019sma}, I define centrality classes from the amount of produced entropy, $S$.
\begin{figure}[t]
    \centering
    \includegraphics[width=.88\linewidth]{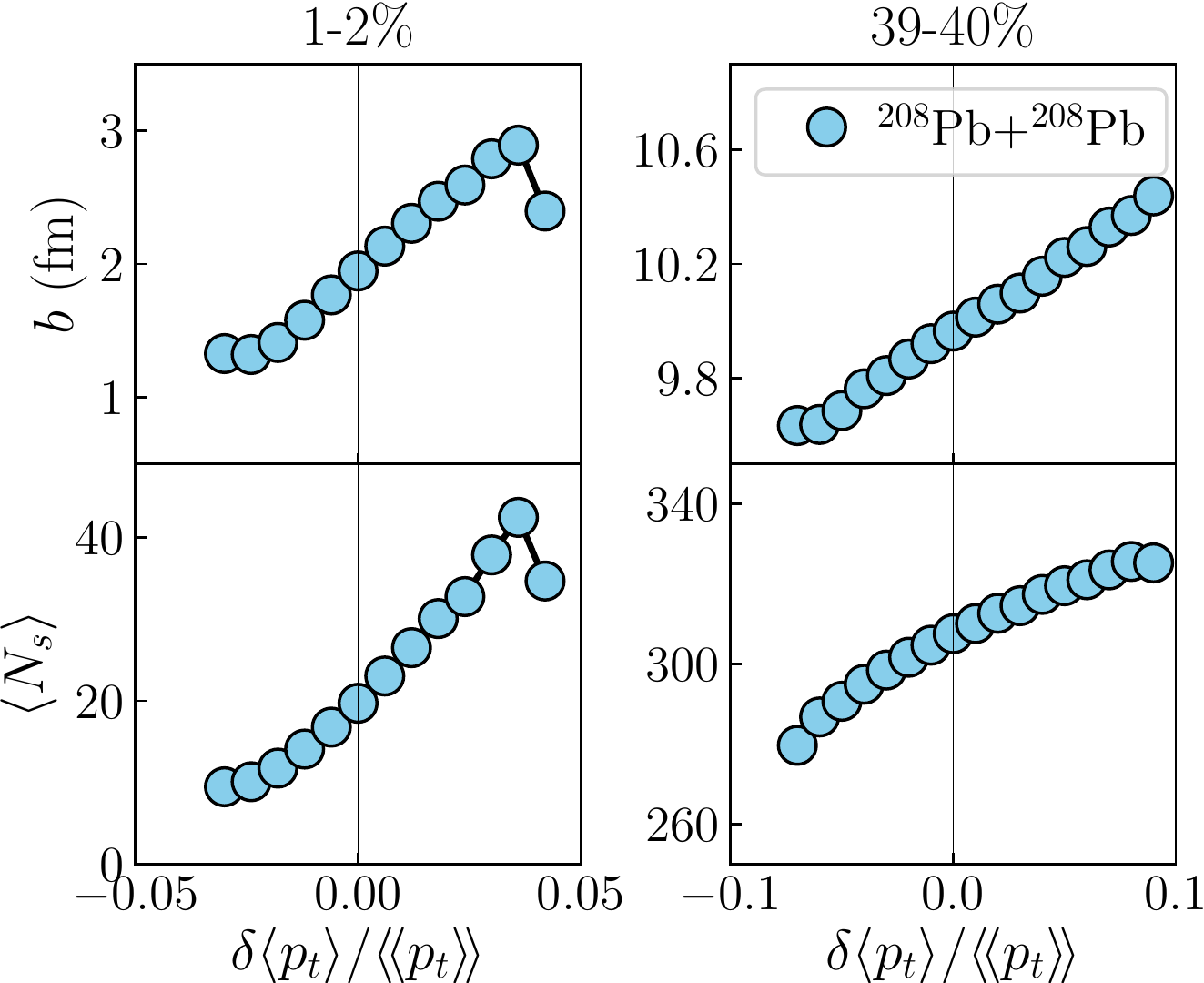}
    \caption{Top: Average impact parameter in $^{208}$Pb+$^{208}$Pb collisions at top LHC energy as a function of $\bra p_t \ket$. Bottom: Average number of spectator nucleons. Left: central collisions, 1-2\%. Right: Peripheral collisions, 39-40\%.}
    \label{fig:1}
\end{figure}

The results of the model can be found in the leftmost panels of Fig.~\ref{fig:1}, where I analyze $^{208}$Pb+$^{208}$Pb collisions with a central cut, 1-2\%. The upper panel shows the collision impact parameter, $b$, as a function of $\bra p_t \ket$. I remark that $b$ and $\bra p_t \ket$ are positively correlated, and that the impact parameter increases by a significant factor from low to high values of $\bra p_t \ket$. The lower panel shows instead the average number of spectator nucleons, $\bra N_s \ket$, as a function of $\bra p_t \ket$. One observes a strong correlation between $\bra p_t \ket$ and $N_s$, as the number of spectators increases by as much as a factor 4 moving towards the high-$\bra p_t \ket$ tail.
\begin{figure}[t]
    \centering
    \includegraphics[width=\linewidth]{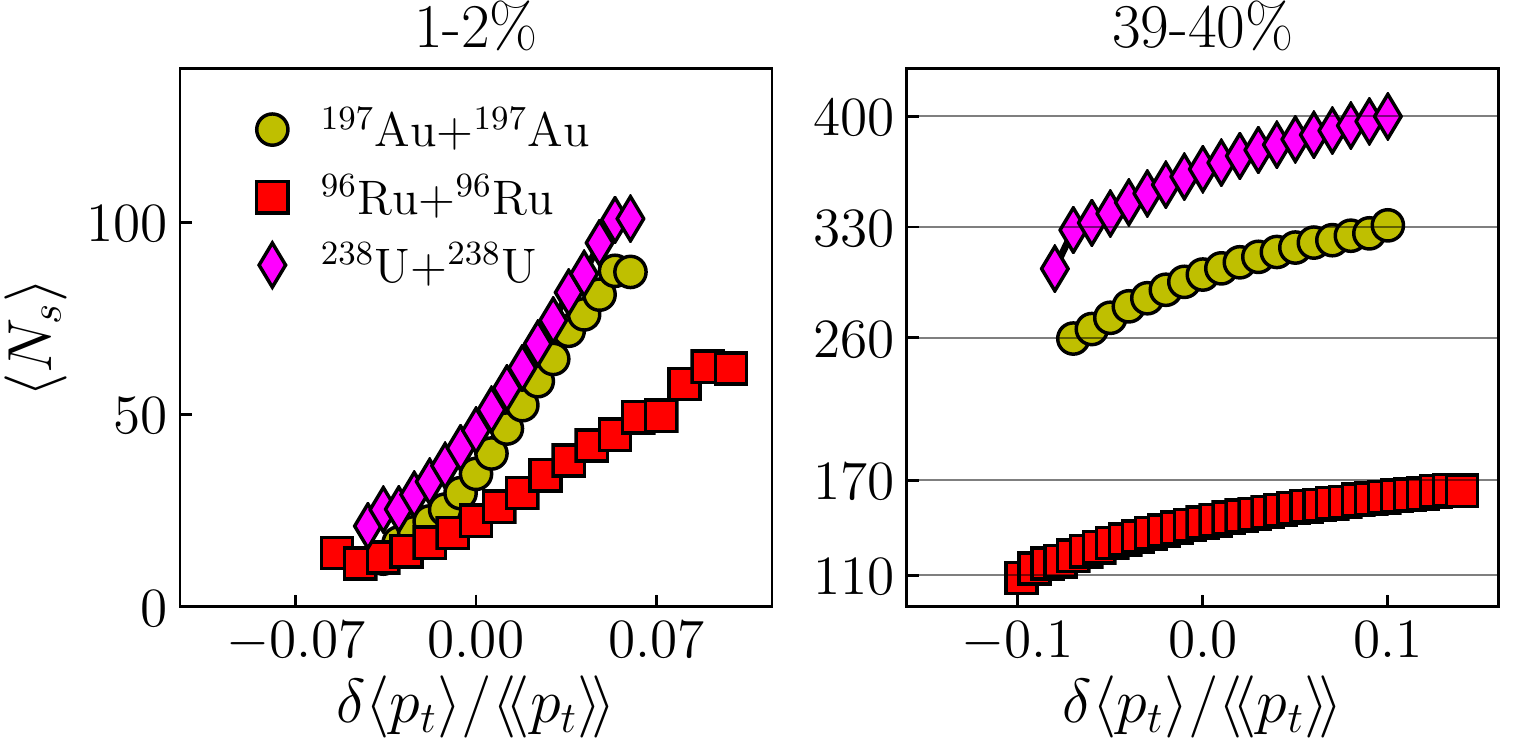}
    \caption{Average number of spectator nucleons in $^{197}$Au+$^{197}$Au, $^{238}$U+$^{238}$U, and $^{96}$Ru+$^{96}$Ru collisions at top RHIC energy. Left: central collisions, 1-2\%. Right: peripheral collisions, 39-40\%.}
    \label{fig:2}
\end{figure}

I assess now the dependence of the previous result on the collision centrality, on the type of colliding species, and on the beam energy. 

The rightmost panels of Fig.~\ref{fig:1} show the same results discussed previously, but in peripheral collisions, corresponding to 39-40\% centrality. One notes that the curves preserve their positive slope, however, the correlation between $\bra N_s \ket$ and $\bra p_t \ket$ is depleted. The overall relative increase of $\bra N_s \ket$ is indeed less than 20\%.

I repeat now the same calculation for systems collided at RHIC. I use the \trento{} model tuned as in Ref.~\cite{Giacalone:2018apa} to simulate $^{197}$Au+$^{197}$Au, $^{238}$U+$^{238}$U collisions, and also $^{96}$Ru+$^{96}$Ru collisions, recently performed at RHIC, whose value of $\kappa_0$ in Eq.~(\ref{eq:predictor}) is chosen by assuming that relative $\bra p_t \ket$ fluctuations scale like $A^{-1/2}$. The left panel of Fig.~\ref{fig:2} shows a very strong correlation between $\bra p_t \ket$ and $\bra N_s \ket$ in central collisions at RHIC. This correlation is in fact stronger in this figure than in the previous one. The reason is that RHIC systems fluctuate more~\cite{Giacalone:2019vwh}, so that to a given collision centrality corresponds a broader range of impact parameters. In central $^{197}$Au+$^{197}$Au, for instance, the number of spectators increases by roughly a factor 7. I further note that the results shown for $^{238}$U+$^{238}$U collisions are obtained by implementing deformed nuclei ($\beta=0.3$). However, while it has been established that a selection of central events based on $\bra p_t \ket$ allows to discern body-body and tip-tip geometries~ \cite{Giacalone:2019pca,Giacalone:2020awm}, this appears to have a negligible impact on the average spectator number, shown in Fig.~\ref{fig:2}. Results for more peripheral collisions are finally shown in the right panel of Fig.~\ref{fig:2}. I note that $\bra N_s \ket$ increases by about 60 units in all systems, an effect which is quite significant in $^{96}$Ru+$^{96}$Ru collisions. One should nevertheless keep in mind that, for peripheral collisions at RHIC beam energy, the physics of $\bra p_t \ket$ is nontrivially influenced by the presence of pre-hydrodynamic flow~\cite{Giacalone:2020byk}, which will have to be properly addressed in future quantitative evaluations.

\begin{figure}[t]
    \centering
    \includegraphics[width=.8\linewidth]{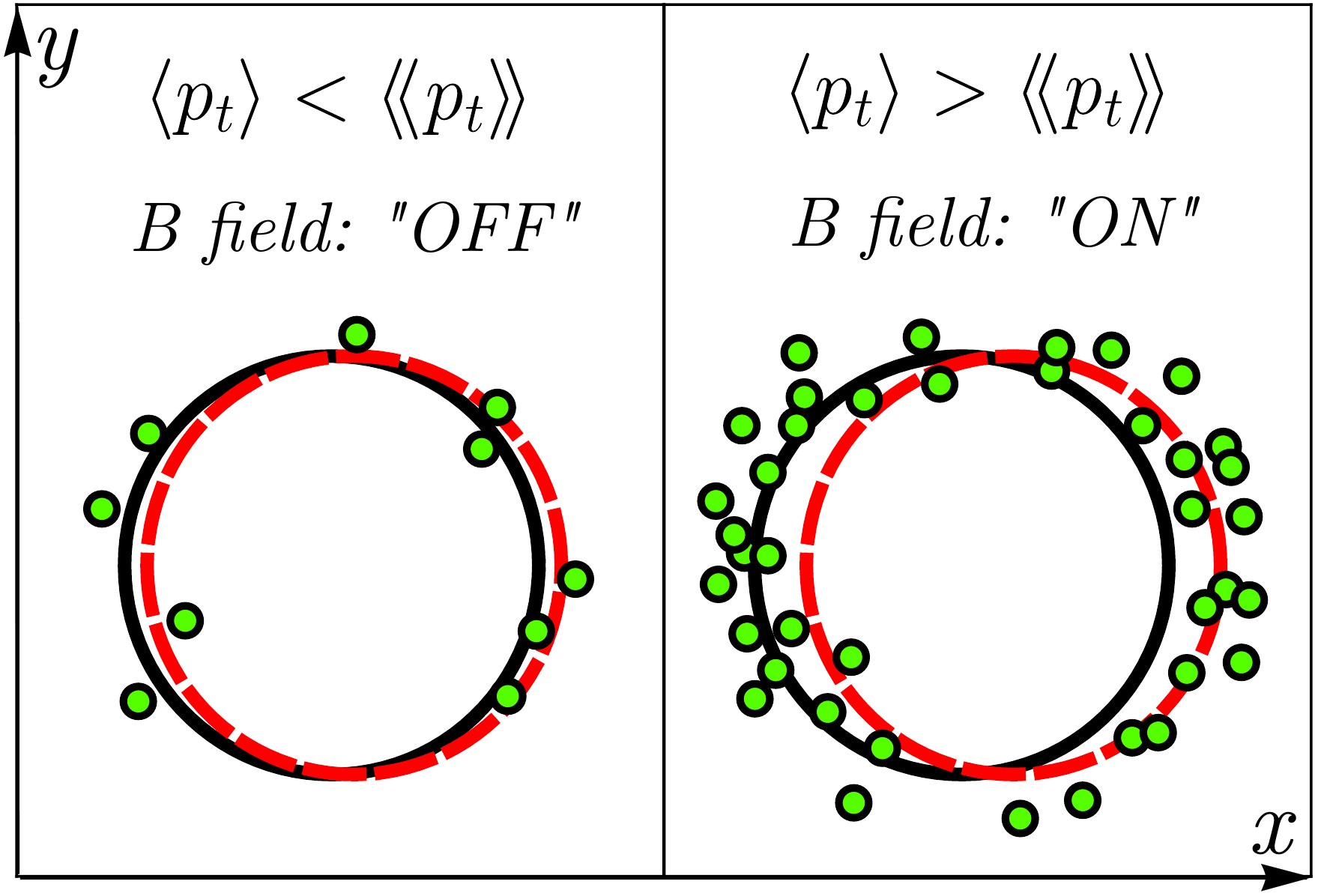}
    \caption{Transverse plane projection of the average geometry of central (1-2\%) collisions of $^{208}$Pb nuclei, and its implication for the manifestation of the $B$ field induced by the spectator protons. Left: low $\bra p_t \ket$, corresponding to $b\sim 1$~fm, and spectator number $N_s=10$. Right: high $\bra p_t \ket$, corresponding to $b\sim2.8$~fm, and $N_s=40$. The spectator nucleons are displayed as small circles.}
    \label{fig:3}
\end{figure}

I summarize these findings in Fig.~\ref{fig:3}, which provides an illustration of my argument for central $^{208}$Pb+$^{208}$Pb collisions. The $B$ field indicated in the figure is the $B$ field induced by the spectators. Collisions at low $\bra p_t \ket$ (left panel in Fig.~\ref{fig:3}) correspond to events at small impact parameter, small number of spectators, and thus a small $B$ field.  Moving to high $\bra p_t \ket$ (right panel in the figure), the impact parameter increases, and this triggers an enhancement in the number of spectator nucleons, which does turn the $B$ field up. The mean transverse momentum, hence, serves as a sort of knob to turn up and down the strong $B$ field created in high-energy nuclear collisions at a given collision centrality. An analogous picture could be drawn for the $B$ field produced by the participant nucleons. 

A couple of comments are in order. The proposed method amounts to an event-shape selection in which one uses $\bra p_t \ket$ to sort events according to their size. Another event-shape engineering method commonly used in heavy-ion analyses uses instead the elliptic flow, $v_2$, to sort events according to their ellipticity, $\varepsilon_2$. In principle, the ellipticity is correlated with the impact parameter, i.e., with the system size, at a given centrality, however, this correlation is small~\cite{Giacalone:2020dln}. Within the \trento{} model, I have indeed checked that the correlation between $\bra p_t \ket$ ad $\bra N_s \ket$ is in fact twice as strong as the correlation between $\varepsilon_2$ and $\bra N_s \ket$. The average transverse momentum stands out, then, as the final-state observable presenting the strongest correlation with the number of spectators (or participants) at a given collision centrality. It is finally important to appreciate that, while the fine details of the results shown in Fig.~\ref{fig:1} and Fig.~\ref{fig:2} depend on the specific \trento{} setup, the fact that the curves have positive slope is fully generic. Hydrodynamics implies only that $\bra p_t \ket$ is proportional to the system size at fixed multiplicity. The larger spectator $B$ field produced at high $\bra p_t \ket$ appears to be, then, a built-in feature of the Glauber modeling~\cite{Miller:2007ri} of nuclear collisions.

This result is remarkably simple, but its phenomenological consequences are vast. The selection of events based on $\bra p_t \ket$ gives a new experimental handle on $N_s$ and $N_{\rm part}$. Therefore, observables and phenomena which are driven by the magnetic should present a nontrivial correlation with the average transverse momentum, a feature which should be investigated in theoretical calculations. One is given an observable, $\mathcal{O}$, and wants to study its correlation with $\bra p_t \ket$. A method to do this, and which allows one to obtain results directly comparable to experimental data, consists in the evaluation a Pearson correlation coefficient, as done, e.g., by Bo\.zek in Ref.~\cite{Bozek:2016yoj}. Dubbing $\delta \bra p_t \ket = \bra p_t \ket - \bbra p_t \kket$, and $\delta\mathcal{O}=\mathcal{O}-\bra \mathcal{O}\ket$, their correlation is defined by:
\begin{equation}
    \rho\left ( \bra p_t \ket,\mathcal{O}  \right ) = \frac{\biggl \bra  \delta \bra p_t \ket \delta\mathcal{O}  \biggr \ket}{\sqrt{ \left \bra \left (\delta \bra p_t \ket \right )^2 \right \ket  \left \bra \left (\delta \mathcal{O} \right )^2 \right \ket}}.
\end{equation}
This quantity isolates the genuine correlation between $\bra p_t \ket$ and $\mathcal{O}$ originating from collective effects. Note that this correlation should be evaluated in a narrow class of multiplicity, e.g., centrality bins of size 1\% or smaller. This is typically doable with experimental data, where millions or billions of collisions are recorded, but can be problematic in full hydrodynamic calculations, due to the limited statistics of events. However, methods to address this issue exist~\cite{Olszewski:2017vyg}, and are currently in use in studies of heavy-ion collisions~\cite{Bozek:2020drh,Schenke:2020uqq}.

I conclude with an example of application of the idea introduced in this manuscript. I explain in particular how it can be applied to the observable used to infer signatures of the chiral magnetic effect (CME) in high-energy nuclear experiments. The CME is a manifestation of local strong parity violation which is expected to occur in relativistic nuclear collisions~\cite{Fukushima:2008xe,Skokov:2016yrj}. At the high temperatures achieved in the early stages of the quark-gluon plasma, one expects the emergence of local domains of chirally-imbalanced matter with a nonzero axial chemical potential, $\mu_5$. In presence of an external magnetic field, such as that produced by the spectator protons, an electric current is thus induced, $\vec J \propto \mu_5 \vec B$. Positively- and negatively-charged particles get pushed (in opposite directions) along this current, i.e., along the direction of the $B$ field~\cite{Kharzeev:2004ey}.  

The CME is thus a dipole-like charge-dependent deformation of the system in momentum space, and as such it contributes to the variance of the charge-dependent hadron dipolar flow, $v_1^\pm\equiv\bra \cos ( \phi_1^\pm - \phi_2^\pm ) \ket$, in the final state. Since in off-central collisions the direction of the $B$ field and the direction of the impact parameter are strongly correlated~\cite{Bloczynski:2012en,Hou:2020zhb}, the signal of the CME is typically measured~\cite{Voloshin:2004vk} as a correlation between the plane of $v_1^\pm$ and the reaction plane, which, up to fluctuations, is the same as the plane of elliptic flow, $v_2$. This corresponds to the following 3-particle correlation:
\begin{equation}
   g^{\pm} = \biggl \bra \cos \left (   \phi_1^\pm + \phi_2^\pm - 2\phi_3 \right)  \biggr \ket = \left (v_1^\pm \right )^2  v_2,
\end{equation}
where $v_2 \equiv \bra \cos 2(\phi_1-\phi_2)\ket$ is the elliptic flow of all hadrons. Now, the strength of the CME signal grows with the strength of the $B$ field, and thus, according to my results, it should increase with $\bra p_t \ket$ at a given collision centrality. The relevant measure of the correlation between the CME signal and $\bra p_t \ket$ is hence given by the following 3-particle correlator:
\begin{equation}
\label{eq:corr}
\rho^\pm \left ( \bra p_t \ket , v_1^\pm \right) =  \frac{\biggl \bra \delta\bra p_t \ket \cos \left ( \phi_1^\pm - \phi_2^\pm \right ) \biggr \ket } { \sqrt{ \left \bra \left (\delta\bra p_t \ket \right )^2   \right \ket  \left (  g^{\pm}/v_2  \right ) }}.
\end{equation}
This gives the statistical correlation between $v_1^\pm$ and $\bra p_t \ket$. In presence of CME signal, this quantity is positive, whereas a baseline for its value in absence of CME could be estimated following the calculations of Ref.~\cite{Schenke:2019ruo}. 

Indeed, like in the case of $g^\pm$, \textit{background} effects~\cite{Wang:2009kd,Schlichting:2010qia,Bzdak:2012ia,Bozek:2017thv} contribute to $\rho^\pm$, although this observable presents novel qualitative features. First of all, major background contributions to the CME signal, such as global momentum conservation~\cite{Bzdak:2012ia}, scale like $1/N$, where $N$ is the multiplicity. The positive slope of the curves shown in Fig.~\ref{fig:1} and in Fig.~\ref{fig:2} occurs, however, at fixed $N$. Hence, the selection of events based on $\bra p_t \ket$ allows one to enhance the CME signal, and to yield a positive value for $\rho^\pm$, while keeping the background fixed. Additionally, background phenomena get themselves correlated with $\bra p_t \ket$ in the evaluation of $\rho^\pm$. Suppose that a given background effect is responsible for a large fraction of the CME signal, $g^\pm$, measured at a given collision centrality. However, if such a background effect is uncorrelated with $\bra p_t \ket$, then it does not contribute to $\rho^\pm$. I thus strongly recommend experimental investigations of $\rho^\pm$, which should pave the way for new studies of CME-related effects, especially in collisions at small impact parameter, that will nicely complement the ongoing searches.

I reiterate that this application concerns only the CME signal. A correlation such as that given by Eq.~(\ref{eq:corr}), with its nontrivial implications, should be indeed constructed for all observables that present a sensitivity to the strong $B$ field produced in high-energy nuclear collisions.

% Acknowledgments
I thank Piotr Bo\.zek, Sandeep Chatterjee, Eduardo Grossi, Niseem Madgy, Jean-Yves Ollitrault, Chun Shen, and Prithwish Tribedy for useful discussions and comments on the manuscript.

\end{document}